\documentclass[12pt,fleqn]{article}
\usepackage{epsfig,times,latexsym,enumerate,lscape}
\newcommand{\prd}{Phys. Rev. D }
\newcommand{\pp}{Preprint gr-qc/}

\begin{document}
\title{Data analysis of continuous gravitational wave: All sky search and study
of templates}
\author{D.C. Srivastava$^{1,2}$\thanks{e-mail: dcsrivastava@now-india.com} 
      $\,$ and S.K. Sahay$^1$\thanks{e-mail: ssahay@iucaa.ernet.in}\thanks{Present address: Inter University Centre for Astronomy and Astrophysics, Post Bag 4, Ganeshkhind, Pune - 411007, India}\\ \\
\normalsize $^1$Department of Physics, DDU Gorakhpur University, Gorakhpur-273009, U.P., India.\\
\normalsize $^2$Visiting Associate, Inter University Centre for Astronomy and Astrophysics, Post  \\
\normalsize Bag 4, Ganeshkhind, Pune-411007, India.}
\date{}
\maketitle
\begin{abstract}
We have studied the problem of all sky search in reference to continuous
gravitational wave particularly for such sources whose wave-form are known
in advance. We have made an analysis of the number of templates required for
matched filter analysis as applicable to these sources. 
We have employed the concept of {\it fitting factor\/} {\it (FF)\/}; treating the 
source location as the parameters of the signal manifold and have studied 
the matching of the signal 
with templates corresponding to different source locations. We have investigated the 
variation of FF with source location and have noticed a symmetry in template 
parameters, $\theta_T$ and $\phi_T$. It has been found that the two different template values in 
source location, each in $\theta_T$ and $\phi_T$, have same {\it FF\/}. We have also 
computed the number of templates required assuming the noise power spectral density
$S_n(f)$ to be flat. It is observed that higher
{\it FF\/} requires exponentially increasing large number of templates.
\end{abstract}

\section{Introduction}
\indent Gravitational wave (GW) Laser Interferometer antennas are essentially omni - 
directional with their response better than 50\% of the average over 75\% of the whole sky
(Grishchuk et al., 2000). Hence the data analysis systems will have to carry
out all sky searches for its sources. We know that the amplitude of intense GW
believed bathing the earth is very small, as compared to the sensitivity of GW
detectors, and is further masked by the dominant noise. In these circumstances,
continuous gravitational wave (CGW) sources are of prime importance because for such sources we can 
achieve
enhanced signal-to-noise ratio (SNR) by investigating longer observation data set. However, a long
observation time introduces modulation effects, arising due to the relative
motion of the detector and the source. 
As a consequence, there results redistribution of power in the forest of side bands 
resulting into the reduction of the expected power due to amplitude modulation (AM). 
The problem of all sky search gains
another dimension in view of the fact that there are reasons to believe the
presence of intense GW sources whose locations and even frequencies are not
known. Amongst such sources pulsars occupy an important position. Similar to
all sky search one will also have to do all frequency search. All sky all
frequency search is the holy grail of gravitation pulsar astronomy. In this
paper we confine ourselves to the problem of all sky search.

\par Search of CGW without a priori knowledge appears to be
computationally quite demanding even by the standard computers
expected to be available in the near future. For example, in the case of
bandwidth $10^3$ Hz, observation time $10^7$ sec. and star's minimum
decay time  of
$100$ years one would require $10^{14}\, Tflops$ computer (Frasca, 2000). Very fast
computer and large memories with ample amount of disk space seems inevitable.
However, a choice of optimal data processing and a clever programming is also
integral part of a solution to this problem. Amongst these the pre-correction
of time series due to the Doppler modulation before the data is processed may be
a method, which will reduce the computational requirements. In reference to this, 
Schutz (1991) has introduced the concept of patch in the sky as the 
region of space throughout which the required Doppler correction remains the same. He
has also demonstrated that the number of patches required for $10^7$ sec.
observation data set and one KHz signal would be about $1.3 \times 10^{13}$. However, 
the size of the patch would also depend on the data analysis technique being
employed. 

\par Matched filtering is the most suitable technique for the detection of signals
from sources viz., pulsars whose wave form is known. The wave forms are used
to construct a bank of templates, which represent the expected signal wave form
with all possible ranges of its parameters. The time of arrival, source
location, frequency of the signal,
ellipticity of the source and its spin down represent important parameters of GW
emitted by a pulsar. For detection of GW we check if the
cross correlation of the templates with the corresponding data set exceeds the
preassigned threshold. We introduce in the next section the criterion of the 
{\it fitting factor (FF) \/} (Apostolatos, 1995) applicable to such analysis. We 
consider the source location as parameters of the signal manifold and investigate the 
matching of the waveforms corresponding to different source locations. In section 3 we 
compute the number of templates required for all sky search. A discussion of the 
results is provided in the section 4. 

\section{Matched filter analysis: Templates}
\label{sec:templates}
The bank of templates is matched, in practice, to only a discrete set of signals from
among the continuum of possible signals. Consequently, it is natural that all the 
signals will not get detected with equal efficiency. However, it is possible to 
choose judiciously the set of templates so that all the signals of a given amplitude are
detected with a given minimum detection loss. The standard measure for
deciding what class of wave form is good enough is the {\it FF\/}.  
It quantitatively describes the closeness of the true signals to the
template manifold in terms of the reduction of SNR arising due to the cross 
correlation of 
a signal outside the manifold with the best matching templates lying inside the 
manifold. If the {\it FF\/} of a template family is unity the signal lies in the 
manifold. If the {\it FF\/} is less than unity the signal lies outside manifold.

\par Even if the signal discrete templates lie within the template manifold
it would be unlikely that any of the actual templates used would correspond to the
signal. The parameters describing the search template
(source location, ellipticity, etc.) can vary continuously through
out a finite range of values. The set of templates characterised by
the continuously varying parameters is of-course infinite. However, in
practice the interferometer output must be cross correlated with a finite 
 subset
of the templates whose parameter values vary in discrete steps from
one template to the next. This subset (``the discrete template
family'') has measure zero on the manifold of the full set of
possible templates (``the continuous template family''), so the
template which most closely match a signal will generally lie in
 between the signal and the nearest of the discrete template
family. The mismatch between the signal and the nearest of the
discrete templates will cause some reduction in SNR. This would mean that
the members of the discrete template
family must be chosen so as to render acceptable loss of SNR.

\par The study of templates has been made by many research workers in time
domain [Schutz
(1991), Kr\'olak (1997), Brady et. al. (1998), Brady and Creighton (2000),
Jaranowski et al. (1998) and Jaranowski and  Kr\'olak (2000)]. However,
the analysis in the frequency domain has the advantage of incorporating
interferometer's spectral noise density. In order to determine the number
of templates required for matched filtering analysis, we make use of the formula for 
{\it FF\/}  as given by Apostolatos (1995) i.e.
\begin{eqnarray}
FF & = &  \max\limits_{\theta , \phi} \frac{\langle h(f)|
h_T(f;\theta_T , \phi_T)\rangle}{\sqrt{\langle h_T(f;\theta_T , \phi_T )|h_T(f;
\theta_T , \phi_T )\rangle\langle h(f)|h(f)\rangle}}
\label{eq:ff1}
\end{eqnarray}

\noindent where $h(f)$ and $h_T(f; \theta_T , \phi_T)$ represent respectively the
FTs of the actual signal wave form and the templates.  

\par In earlier papers we have made an analysis of continuous gravitational
wave output through Fourier transform (FT) (Srivastava and Sahay, 2002 a, b).
These papers are referred in the sequel respectively as I and II. It has been 
established there that the AM of
CGW data output results into redistribution of power at four additional frequencies
$f \pm 2f_{rot}$, $f \pm f_{rot}$ in accordance with the frequency modulation (FM).
Hence it is sufficient for the analysis of {\it FF\/} to consider only the
frequency modulated FT. The results obtained in paper-II regarding FT
of frequency modulated data output i.e. (II-27, 28) may be arranged, making use of  
the symmetry property of the Bessel functions, as follows. 
\begin{eqnarray}
\tilde{h}(f)&\simeq & { \nu \over w_{rot}}\left[ {J_o({\cal Z}) J_o
({\cal N}) \over 2\nu^2}\left[ \{ \sin ( {\cal R} + {\cal Q} ) - \sin
({\cal R} + {\cal Q} - \nu\xi_o )\}\; + \right. \right. \nonumber\\ &&
i \left. \{ \cos ( {\cal R} + {\cal Q} ) - \cos ({\cal R} + {\cal Q} - \nu
\xi_o )\} \right]\; + \nonumber \\
&& J_o ({\cal Z})\sum_{m = 1}^{m = \infty} {J_m({\cal N})\over 
\nu^2 - m^2} \left[ ( {\cal Y} {\cal U} -  {\cal X} {\cal V} ) - i ( 
{\cal X} {\cal U} + {\cal Y} {\cal V} ) \right]\; + \nonumber \\ 
&& \left.\sum_{k = 1 }^{k = \infty}\sum_{m = -
\infty}^{m = \infty} e^{ i {\cal A}}{\cal B}\left(
\tilde{{\cal C}} - i\tilde{{\cal D}} \right)\right]\; ;
\label{eq:fm_code}
\end{eqnarray} 

\begin{equation}
\left.\begin{array}{ccl}
{\cal X}& =& \sin ({\cal R}  + {\cal Q} - m \pi/2 )\\
{\cal Y}& =& \cos ({\cal R} +{\cal Q} - m \pi/2 )\\
{\cal U}& =& \sin \nu\xi_o \cos m ( \xi_o - \delta ) - {m\over \nu}\left\{\cos
\nu\xi_o \sin m ( \xi_o - \delta ) - \sin m\delta\right\}\\
{\cal V}& =& \cos \nu\xi_o \cos m ( \xi_o - \delta ) + {m\over \nu}\sin 
\nu\xi_o \sin m ( \xi_o - \delta ) - \cos m\delta\\
\end{array}\right\}
\end{equation}

\begin{figure}
\centering
\epsfig{file=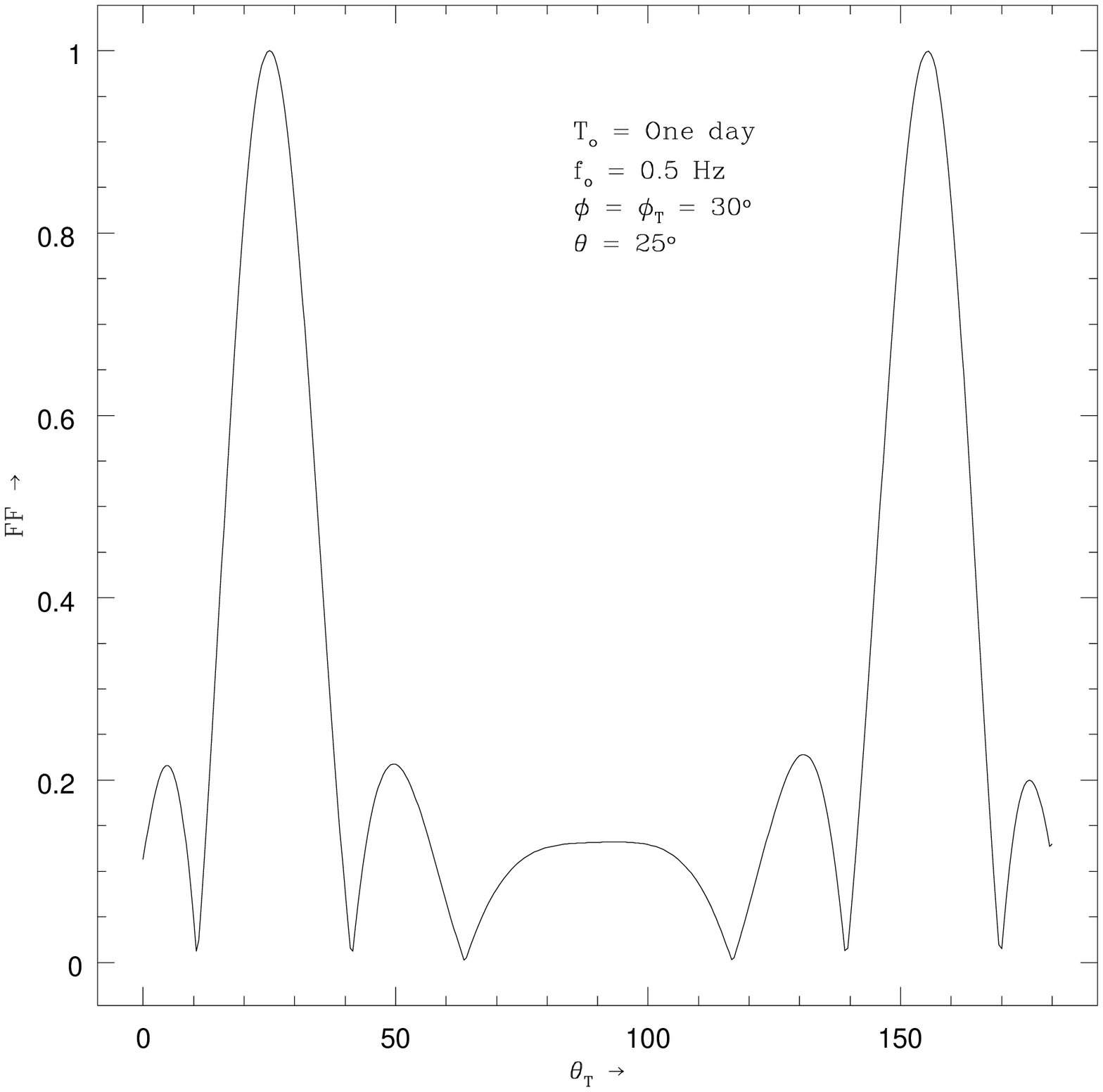,height=16.0cm}
\caption{ Variation of $FF$ with $\theta_T$.}
\label{fig:symtheta}
\end{figure}

\begin{figure}
\centering
\epsfig{file=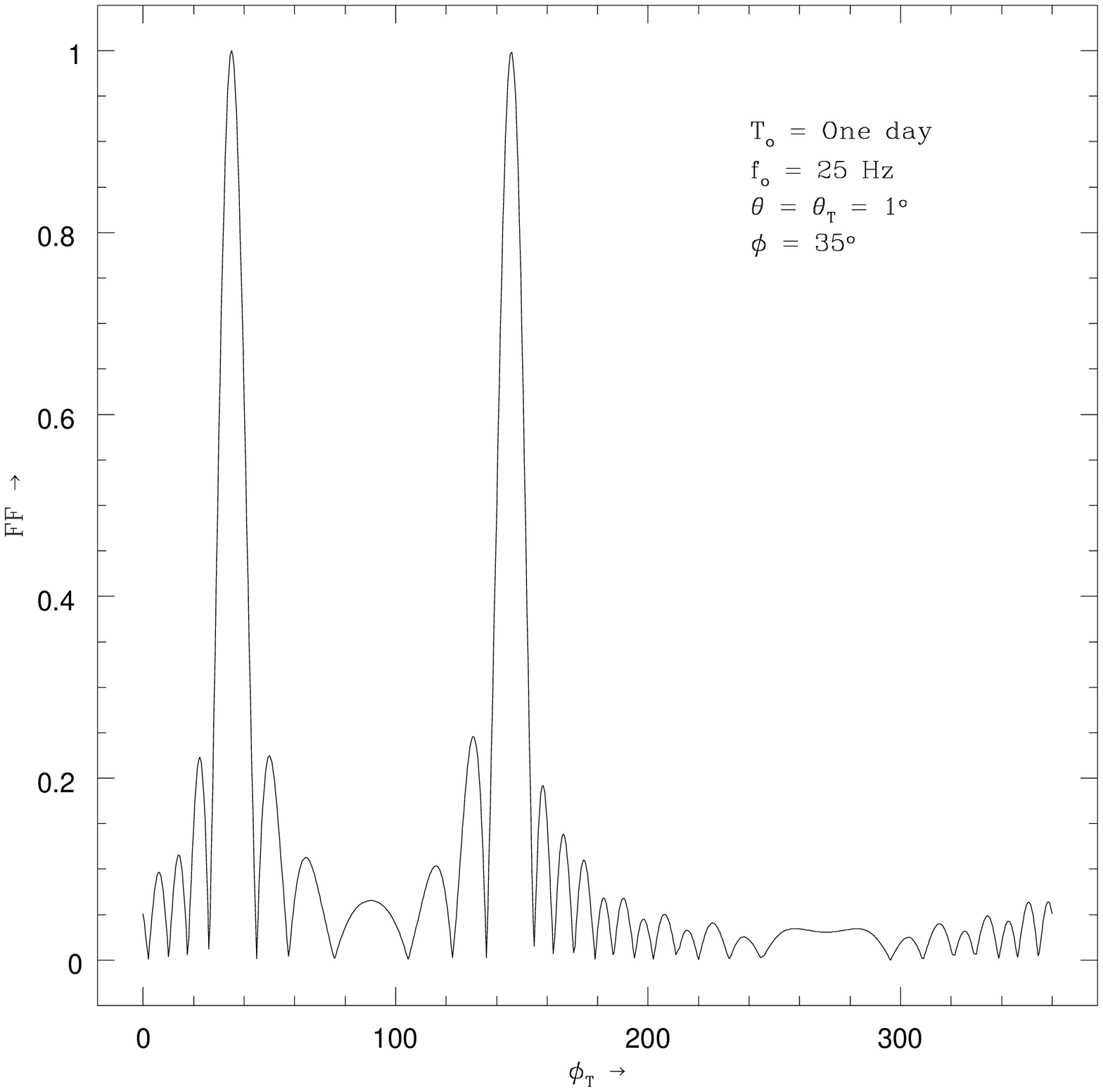,height=16.0cm}
\caption{Variation of $FF$ with $\phi_T$.}
\label{fig:symphi35}
\end{figure}

\begin{figure}
\centering
\epsfig{file=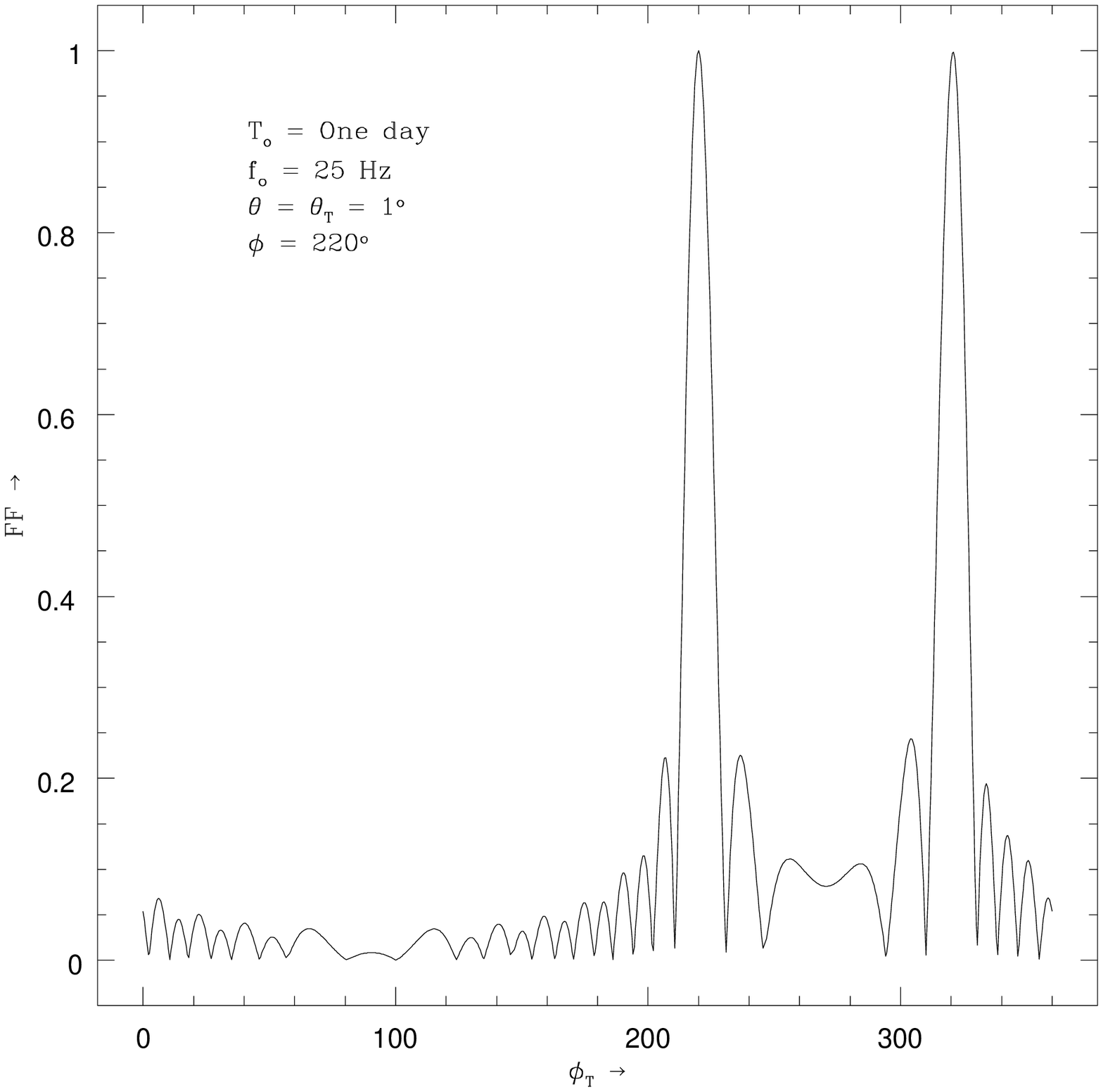,height=16.0cm}
\caption{Variation of $FF$ with $\phi_T$.}
\label{fig:symphi220}
\end{figure}

\noindent Now it is straight-forward to compute {\it FF\/}. To understand the
procedure let us consider a source at location $(\theta , \phi ) = (25^o , 30^o)$ 
emitting frequency $f_o = 0.5 $ Hz. In order to evaluate the summation given in 
Eq.~(\ref{eq:fm_code}) let us note that the value of Bessel function decreases rapidly 
as its order exceed the argument. Accordingly, for the present case it is sufficient 
to take the ranges of $k$ and 
$m$ as $1$ to $1610$ and $- 4$ to $4$ respectively. We 
wish to analyze the data set for $T_o =$ one sidereal day. 
To maximize {\it FF\/} over $\theta$ and $\phi$, we first maximize
over $\phi$ by fixing $\phi_T$ to some arbitrarily selected value say,
$\phi_T = \phi = 30^o$. Having done this we maximize over $\theta$ 
by varying $\theta_T$ in discrete steps over its entire range
i.e. $0^o$ to $180^o$. The results obtained are plotted in
Fig.~(\ref{fig:symtheta}).
It is remarked that in order to compute the inner product of two waveform
$h_1$ and $h_2$ which is defined as
\begin{eqnarray}
\langle h_1|h_2\rangle & =& 2\int_0^\infty \frac{\tilde{h}_1^*(f)\tilde{h}_2(f)
+ \tilde{h}_1(f)\tilde{h}_2^*(f)}{S_n(f)}df \nonumber \\
 & = &
4\int_0^\infty \frac{\tilde{h}_1^*(f)\tilde{h}_2(f)}{S_n(f)}df
\label{eq:ip}
\end{eqnarray}

\noindent where $^*$ denotes complex conjugation, $\tilde{}$ denotes the Fourier 
transform of the quantity underneath 
$(\tilde{a}(f) = \int_\infty^\infty a(t)exp(- 2 \pi i f t)dt) $ and $S_n(f)$ is the 
spectral density of the detector's noise. One would require to integrate the 
expression over the band width of Doppler modulated signal. The band width may be
determined by computing the maximum value of the Doppler shift. In accordance with
Eq. (I-23) the Doppler shift is $\sim 10^{-4} f_o$. However, for the present case we 
consider the band width equal to $0.002$ Hz. In a similar manner one may fix the 
$\theta_T$ and obtain the variation of {\it FF\/} with template
parameter $\phi_T$. Figures~(\ref{fig:symphi35}) and~(\ref{fig:symphi220})
represent such a plot for a signal of $f_o = 25$ Hz emitted by a 
source located at $\theta = 1^o$ and  $\phi = 35^o$ and $220^o$, respectively.\\

\noindent The following points in reference to these plots may be noted.
\begin{enumerate}[(i)]

\item The {\it FF\/} is unity for $\theta_T = 25^o$, $155^o$ [Fig~(\ref{fig:symtheta}]
for $\phi_T = 35^o$, $145^o$ [Fig~(\ref{fig:symphi35})] and for $\phi_T = 220^o$,
$320^o$ [Fig~(\ref{fig:symphi220})].

\item It is found that the dependence of {\it FF \/} on the 
template variables $\theta_T$ and $\phi_T$ may be expressed via  
\begin{equation}
FF = e^{- 0.00788(\theta \; - \;\theta_T)^2}
\label{eq:thetaff}
\end{equation}
\begin{equation}
FF = e^{- 0.01778(\phi\; - \;\phi_T)^2}
\label{eq:phiff}
\end{equation}

\item The oscillatory behaviour is more or less typical in waveform that match
well or bad depending on their parameters. However, we are content with the
technique we have employed as the region
of such artificial facets fall in the region of the $FF < 0.25$.
\end{enumerate}

\noindent Finally, we conclude this section by noting the symmetry property of the 
{\it FF\/} with template parameters. A closer look of the graphs and the remark (i)
above reveal the following symmetry property. The $FF$ is symmetrical under 
following transformations.
\begin{equation}
\theta_T \longrightarrow  \pi - \theta_T  \qquad 0 \le \theta_T \le \pi
\end{equation}
\begin{equation}
\phi_T \longrightarrow \pi - \phi_T \qquad 0 \le \phi_T \le \pi
\end{equation}
\begin{equation}
\phi_T \longrightarrow 3\pi - \phi_T \qquad \pi \le \phi_T \le 2\pi 
\end{equation}

\noindent Let us note that these symmetry properties are based on our results
for one day observation time. The generic nature of the symmetries
may be established only after studying the variation of $FF$ with $T_o$.
Unfortunately, we could not make this analysis because of our limitations on the 
computational facilities. 

\section{Number of Templates}
It is important to study the problem of number of templates for all sky
search in the light of $FF$.
The results of the previous section reveal that the grid
spacing $\bigtriangleup \theta$ in the $\theta-$parameter of templates may be
expressed symbolically as a function of $FF$, $f_o$, $T_o$, $\theta$ and $\phi$ as   
\begin{equation}
\bigtriangleup\theta = {\cal F}(FF, f_o, \theta ,\phi ,T_o )
\end{equation}
\noindent Similarly, we have
\begin{equation}
\bigtriangleup\phi = {\cal G}(FF, f_o, \theta ,\phi ,T_o)
\end{equation}

\par In the literature it is reported that the choice of 
grid spacing $\bigtriangleup\phi$ depends insignificantly on $\phi$, whereas 
$\bigtriangleup \theta$ depends as $\sin 2\theta$ (Brady and Creighton, 2000). 
However, in order to arrive at such conclusions one may have to make analysis for 
different values of $\theta$ and $\phi$. Unfortunately, due to the limited memory and 
the efficiency of the computer we could not make a study of this aspect.

\noindent In view of this, Eqs.~(\ref{eq:thetaff}) and~(\ref{eq:phiff}) may be equivalently expressed as
\begin{equation}
{\cal F}(FF, 0.5, 25^o , 30^o ,T) = \left[ - (0.00788)^{-1}\ln (FF)\right]^{1/2} 
\end{equation}
\begin{equation}
{\cal G}(FF, 25, 1^o , 35^o ,T) = \left[ - (0.01778)^{-1}\ln (FF)\right]^{1/2} 
\end{equation}

\noindent For any chosen value of $FF$ one can determine $\bigtriangleup\theta$ and
$\bigtriangleup\phi$. However,  there is no unique choice for it. 
Our interest would be in the assignment of
$\bigtriangleup\theta$ and  $\bigtriangleup\phi$ such that the 
spacing is maximum resulting into the least number of
templates. As we have mentioned earlier,  
there is stringent requirement on reducing computer time. Accordingly, 
there is serious
need of adopting some procedure/formalism to achieve this. For example, one
may adopt the method of
hierarchical search given by Mohanty and Dhurandhar (1996) and 
Mohanty (1998). This
search is carried out in two steps. At the first level one would start with
template bank with a coarse spacing in the
parameter space but with a lower threshold. At the next level a more
finely spaced set of templates and a higher threshold would be used
but only around those templates of the previous level which exceeded the
previous threshold.

\par
\begin{figure}
\centering
\epsfig{file=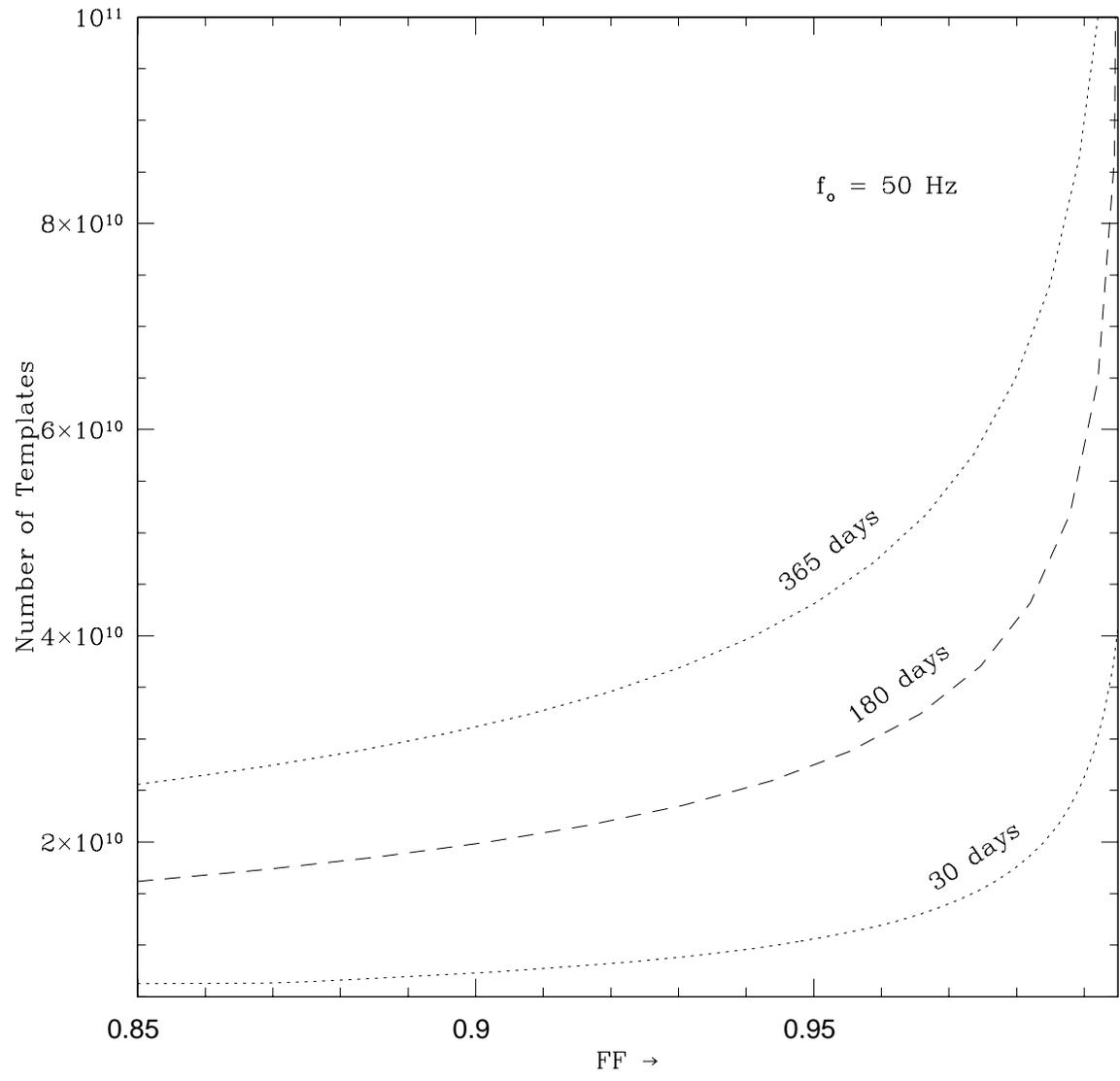,height=16.0cm}
\caption{Variation of number of templates with $FF$ for fixed $f_o$ at
different $T_o$.}
\label{fig:temtime}
\end{figure}

\begin{figure}
\centering
\epsfig{file=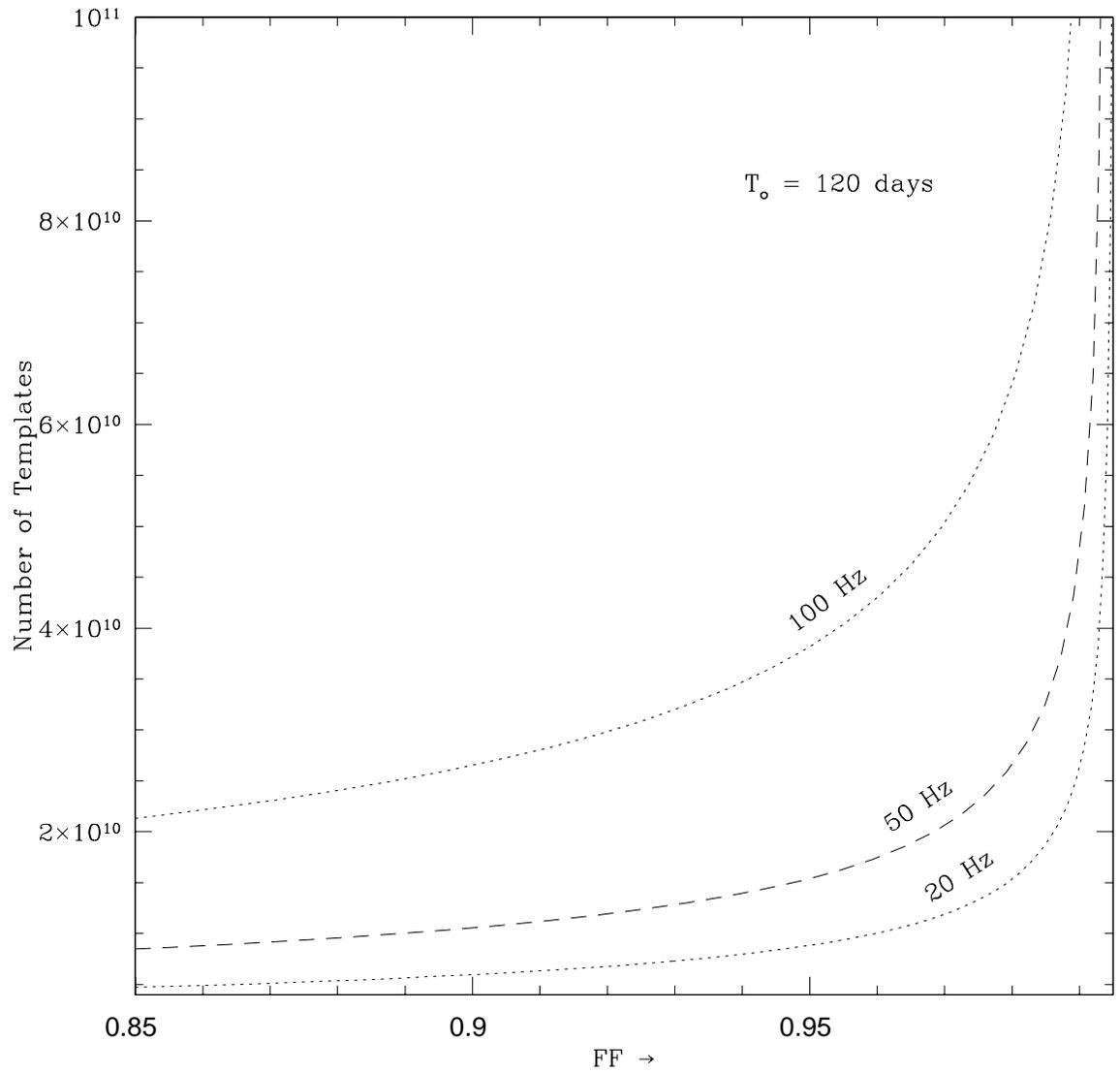,height=16.0cm}
\caption{Variation of number of templates with $FF$ for fixed $T_o$ and of
frequencies.}
\label{fig:temfreq}
\end{figure}

 However, an important issue related to the problem of number of templates
is regarding the study of the 
behaviour of number of templates with $FF$ for different
$f_o$ and $T_o$. We have made an investigation of this aspect. We assume a source
location $(\theta , \phi ) = (1^o , 30^o )$. We choose some value of $FF$, say
$0.995$. Now we maximize $FF$ over $\phi$ by taking $\phi = \phi_T = 30^o$ 
and choose the spacing $\bigtriangleup\theta$ so as to yield the selected $FF$. 
In the case under investigation $\bigtriangleup\theta$ is found to equal 
$4.5 \times 10^{-5}$. Thereafter we maximize over $\phi$ by introducing 
spacing $\bigtriangleup\phi$ in the so obtained bank of templates and determine 
the resulting $FF$. The results obtained may be expressed in the form of a graph 
such as shown in
Figs.~(\ref{fig:temtime}) and~(\ref{fig:temfreq}). Interestingly, the nature
of these curves are similar. We have obtained a best fit to the graphs as
\begin{equation}
N_{Templates} = \exp [a -bx + cx^2 -dx^3 + ex^4]\; ;\qquad 0.85 \le x \le 0.99
\end{equation}

\noindent where $a$, $b$, $c$, $d$ and $e$ are constants as given in 
Table~(\ref{table:coefficients}). 

\begin{table}
\begin{tabular}{|c|c|c|c|c|c|c|}
\hline
&&&&&&\\
$f_o$ & $T_o$ & a & b & c & d & e\\
(Hz)&(days)&$\times 10^{- 2}$&$\times 10^{- 4}$&$\times 10^{- 5}$&$\times 10^{- 6}$&$\times 10^{- 7}$\\
\hline
&&&&&&\\
& 30 & $2138.05 $ & $2071.43$ & $7225.73 $
& $6239.43$ & $2036.14$\\ 
&&&&&&\\ \cline{2-7}
&&&&&&\\
50 & 180 & $2317.05 $ & $- 71.3155$ & $1746.61 $
&$ 2146.55$  &$ 944.931$\\ 
&&&&&&\\ \cline{2-7}
&&&&&&\\
& 365 &$ 2382.96 $ & $216.917 $ &$ 2464.42 $
&$ 2464.42$ & $1031.54$\\ 
&&&&&&\\
\hline
&&&&&&\\
20 & & $2047.55$ & $- 794.473 $ &$ 3564.56 $
&$ 4650.68$ &$ 1945.87$\\ 
&&&&&&\\ \cline{1-1} \cline{3-7}
&&&&&&\\
50 & 120 & $2266.59$ & $4269.44$ & $15655.0$
&$ 17509.5$ & $6484.51$\\
&&&&&&\\ \cline{1-1} \cline{3-7}
&&&&&&\\
100& &$ 2360.23$ & $- 206.906$ & $1158.27$
&$ 1491.01$ & $733.520 $\\ 
&&&&&&\\
\hline
\end{tabular}
\caption{Coefficients of the best fit graphs obtained for the 
number of templates}
\label{table:coefficients}
\end{table}

\par Let us note from the graphs, for sake of comparison, that the number of
templates required for $FF$ equal to $0.97$ are respectively $1.44 \times
10^{10}$, $3.5 \times 10^{10}$ and $5.5 \times
10^{10}$ for observation data sets of 30, 120 
and 365 days and $f_o = 50$ Hz. Similarly to analyse the observation data set
of 120 days of GW frequencies 20, 50 and 100 Hz the number of templates
required are  $1.22 \times 10^{10}$,
$2.16 \times 10^{10}$ and $5 \times 10^{10}$. It is observed that higher
$FF$ requires exponentially increasing large number of templates.

\section{Discussion}
\label{sec:concl5}
In view of the complexity of the FT, which
contains trignometric as well as Bessel functions; one has to be careful in
computing $FF$. We have found useful to employ the 
Romberg integration using Pad\'e approximation. We have used
(i) QROMO of numerical recipes instead of QROMB as the former takes care of
singularities, and (ii) RATINT routine for Pad\'e approximation.

\par We  have noticed marked symmetries in all sky search in both $\theta$ and
$\phi$ space for one day observation time. It has been found that the two different 
template values in source location, each in $\theta_T$ and $\phi_T$, 
have same {\it FF\/}. Accordingly, computation burden will be reduced by a factor of
four. However, it is not clear whether this symmetry
property can be established analytically as well. The source
location, because of these symmetries is uncertain and some other 
analysis is to be adopted for getting the exact location. We have computed the number 
of templates assuming the noise power spectral density $S_n(f)$ to be flat which is 
justified as the bandwidth is extremely narrow. 

\par The issues of optimum template parameterization and placement, and the
related computational burden have been discussed in the literature by several
authors notably by Sathyaprakash and Dhurandhar (1991), Dhurandhar and Sathyaprakash
(1994), Owen (1996), Apostolatos (1995, 1996), Mohanty and 
Dhurandhar (1996), Mohanty (1998), Owen and Sathayaprakash
(1999). The question of possible efficient interpolated representation
of the correlators is a problem of current interest and remains still unsolved.

\section*{Acknowledgments}
The authors are thankful to Prof. S. Dhurandhar, IUCAA, Pune for stimulating
discussions. The authors are thankful to the anonymous referee 
for making the detailed suggestions which resulted in the improvement of the paper.
The authors are also thankful to IUCAA for providing hospitality 
and computational facility where major part of the work is carried out. This work is supported through
research scheme vide grant number SP/S2/0-15/93 by DST, New Delhi.

\end{document}